\documentclass[aps,prc,twocolumn,superscriptaddress,showpacs,floatfix]{revtex4}
\usepackage{graphicx,color}
\usepackage{braket}
\usepackage{amsmath}
\usepackage{amssymb}

\def\scr{\rm\scriptscriptstyle }
\def\sc{\scriptscriptstyle }

\begin{document}
\title{Semiclassical calculations of complete and incomplete fusion in collisions of weakly bound nuclei}
%
%
\author{G.D. Kolinger}
            \affiliation{ Instituto de F\'\i sica, Universidade Federal do Rio Grande do Sul, Av. Bento Gon\c calves 9500, CP 15051, 91501-970, Porto Alegre, Brazil }
\author{L.F. Canto}
            \affiliation{Instituto de F\'{\i}sica, Universidade Federal do Rio de Janeiro, C.P. 68528, 21941-972 Rio de Janeiro, Brazil}
            \affiliation{Instituto de F\'{\i}sica, Universidade Federal Fluminense, Av. Litoranea s/n, Gragoat\'{a}, Niter\'{o}i, R.J., 24210-340, Brazil}
\author{R. Donangelo}
            \affiliation{Instituto de F\'{\i}sica, Universidade Federal do Rio de Janeiro, C.P. 68528, 21941-972 Rio de Janeiro, Brazil}
            \affiliation{Instituto de F\'{\i}sica, Facultad de Ingenier\'{\i}a, Universidad de la Rep\'ublica, C.C. 30, 11000 Montevideo, Uruguay}
\author{S.R. Souza}
            \affiliation{Instituto de F\'{\i}sica, Universidade Federal do Rio de Janeiro, C.P. 68528, 21941-972 Rio de Janeiro, Brazil}
            \affiliation{Instituto de F\'\i sica, Universidade Federal da Bahia, Campus Universit\'ario de Ondina, 40210-340, Salvador, Brazil}
            \affiliation{Departamento de F\'\i sica, ICEx, Universidade Federal Minas Gerais, Av. Ant\^onio Carlos 6627, 31270-901 Belo Horizonte, Brazil}
\date{today}
\begin{abstract}
We use an improved version of the semiclassical method described in Refs.~\cite{MCD02,MCD08,MCD14}  to evaluate fusion cross sections in collisions of weakly bound 
nuclei. This version takes into account the static effects of the low breakup threshold, uses better bin states in the discretization of the continuum and avoids the excitation
of closed channels. The population of these channels is a consequence of the violation of energy conservation, which is inherent in the semiclassical method. The method 
is employed to evaluate complete fusion and total fusion cross section in collisions of the  weakly bound $^{6,7}$Li projectiles with $^{159}$Tb and $^{197}$Au targets, for
which data is available. The overall agreement between theory and experiment is fairly good.

\end{abstract}



\maketitle

\section{Introduction} \label{sec:intro}

The breakup threshold of typical nuclei is of several MeV. For this reason, the influence of the breakup channel on nucleus-nucleus
collisions at near-barrier energies is negligible. A different situation is encountered for the light nuclei $^6$Li, $^9$Be and $^7$Li, 
which have low breakup thresholds ($B=1.47, 1.57$ and 2.47 MeV, respectively). In collisions involving these nuclei, the breakup 
cross sections are very high and the breakup channel affects elastic scattering and fusion reactions. The influence of the breakup 
channel may be even larger in reactions induced by radioactive beams, which became available at several facilities over the last 
two decades~\cite{BNV13}. In such cases, the separation energy of one or a few nucleons may be less than 1 MeV. Then, the 
weakly bound nucleons may occupy orbits with large radii and the nuclear density presents a halo. A few examples are $^{11}$Be 
($B= 0.50$ MeV), $^8$B ($B= 0.14$  MeV) and $^{11}$Li ($B= 0.37$ MeV), which exhibit respectively a one-neutron, a one-proton
and a two-neutron halo.

\bigskip

In collisions of weakly bound nuclei the fusion cross section is affected in two ways~\cite{CCC03}. First, the 
Coulomb barrier tends to be lower owing to the longer tail of the nuclear density. This {\it static} effect of the low breakup 
threshold enhances fusion at low collision energies. There are also {\it dynamic} effects, resulting from the couplings with the 
breakup channels. Breakup couplings give rise to new fusion processes. First, there is the usual fusion reaction, that also takes 
place in collisions of tightly bound nuclei. In this case, the whole projectile fuses with the target, without undergoing breakup. This is called 
{\it direct complete fusion} (DCF). Then, there are fusion processes following breakup. There is the possibility that one or more, 
but not all, fragments of the projectile fuses with the target. This is called {\it incomplete fusion} (ICF). Different ICF processes can
take place, depending on the fragment (fragments) absorbed by the target. Another possibility is that all fragments of the projectile are 
absorbed sequentially by the target. This process, known as {\it sequential complete fusion} (SCF), leads to the same compound 
nucleus as the DCF. For this reason, no experiment can distinguish between SCF and DCF. A few 
experiments can determine the {\it complete fusion} (CF) cross section, which is the sum {${\rm CF=DQF+SCF}$. This is possible
for a few selected projectile-target combinations (for a review see \cite{CGD15,KGA16,CGD06} and references therein).
However, most experiments can only determine the cross section for {\it total fusion} (TF), which is the inclusive process 
${\rm TF=CF+ICF}$. We remark that, from the experimental point of view, CF and ICF are defined in terms of charges. Namely, CF 
corresponds to the situation where the whole charge of the projectile is absorbed by the target. Otherwise, the fusion reaction is classified 
as ICF. Finally, there is the possibility that the projectile breaks up but none of its fragments fuses with the target. This process is known 
as {\it non-capture breakup} (NCBU). It includes both elastic breakup (EBU) and inelastic breakup, where the target remains in its ground state
and goes to an excited state, respectively.\\

From the theoretical point of view, there are also great difficulties to distinguish CF from ICF. The most powerful theoretical tool to 
describe collisions of weakly bound projectiles is the {\it continuum discretized coupled channel} (CDCC)~\cite{KYI86} approximation, 
which has been introduced to describe the breakup of the deuteron. This method has the nice features of allowing a full quantum 
mechanics description of the breakup process. Hagino {\it et al.}~\cite{HVD00} and Diaz-Torres and Thompson~\cite{DiT02} have 
been able to evaluate individual CF and ICF cross section in the $^{11}$Be + $^{208}$Pb collision, where the projectile breaks up 
as $^{11}{\rm Be} \rightarrow\, ^{10}{\rm Be} + n$. Unfortunately their method to evaluate CF and ICF cross sections cannot be 
extended to collisions of projectiles that break up into fragments of comparable masses.   \\

It is much easier to evaluate individual CF and ICF cross sections in classical and semiclassical time dependent approaches.
These cross sections have been evaluated in Refs.~\cite{HDH04,DHH02}, where both the projectile-target motion and the dynamics 
of the breakup process are described by classical mechanics. Diaz-Torres~{\it et al.}~\cite{DHT07,Dia10,Dia11} developed an 
improved model keeping the concept of a classical trajectory, but treating the breakup of the projectile as a stochastic process. 
Marta {\it et al.}~\cite{MCD02,MCD08, MCD14} went one step further, describing the collision by a semiclassical model. In this 
approach, the projectile-target relative motion is treated by classical mechanics, whereas the breakup of the projectile is 
handled by time-dependent quantum mechanics. \\

The purpose of the present work is to develop further the semiclassical approach of Marta, Canto and Donangelo~\cite{MCD02,MCD08,
MCD14}, improving some approximations of the model and applying it to new collisions. This paper is organised as follows. In Sec. II we
present a detailed description of our semiclassical approach to collisions of weakly bound nuclei. In Sec. III, the method
is applied to collisions of $^{6,7}$Li projectiles on $^{159}$Tb and $^{197}$Au targets. We evaluate CF and TF cross sections
and compare our results with the data of Refs.~\cite{PMB11,MSP06,PTN14}. In Sec. IV we summarize the conclusions of our work.
Some technical details about the evaluation of matrix-elements of the interaction are presented in the appendix.

\section{Theoretical framework} \label{sec:model}

We consider the collision  of a weakly bound projectile formed by two clusters on a heavy target, as in  Refs. \cite{MCD14,MCD08,MCD02}.
The main features of our approach are presented below.\\

\begin{figure}
\includegraphics[scale=0.5]{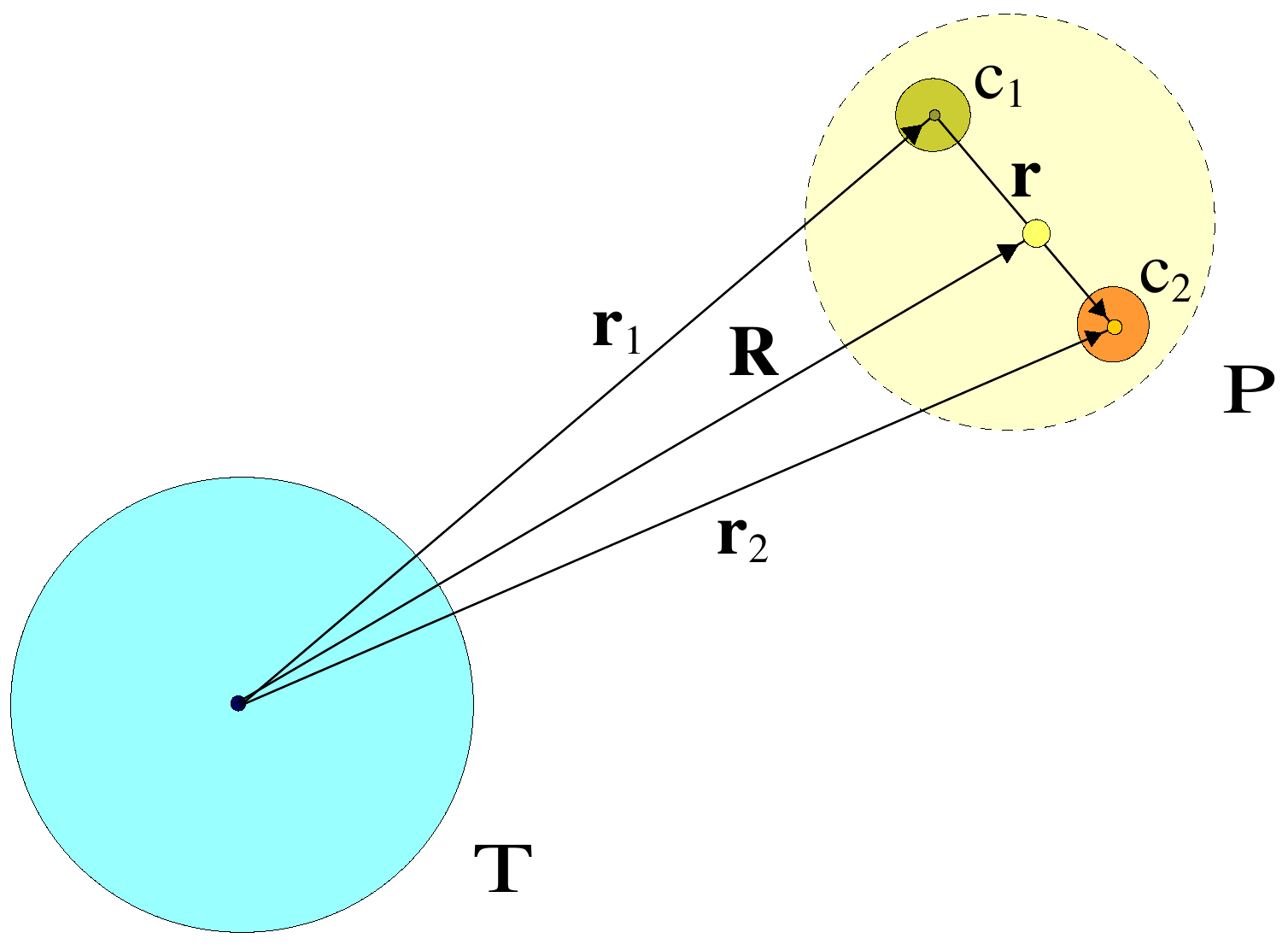}
\caption{Variables used to describe the system configuration.}
\label{fig:system}
\end{figure}

The dynamics of the relative motion between the projectile (with mass and atomic numbers $A_{\scr P}$ and $Z_{\scr P}$) and the 
target (with mass and atomic numbers $A_{\scr T}$ and $Z_{\scr T}$) is described by standard classical equations. More specifically, 
the interaction between the target and  cluster 1 (2), henceforth labeled $c_{\scr 1}$ ($c_{\scr 2}$), is represented by 
$V_{c_{\scr 1}-{\scr T}}({\bf r}_{\scr 1})$  ($V_{c_{\scr 2}-{\scr T}}({\bf r}_{\scr 2})$). 
The vectors ${\bf r}_{\scr 1}$ and ${\bf r}_{\scr 2}$ denote the separation between the centers 
of mass of these fragments and the center of the target, as illustrated in Fig.~\ref{fig:system}. 
These vectors  can be expressed in terms of the projectile-target separation, ${\bf R}$, and the 
vector joining the two clusters, ${\bf r}$, by the usual relations
\begin{equation}
{\bf r}_{\scr 1}= {\bf R}+\frac{A_{\scr 2}}{A_{\scr P}}\ {\bf r}\ \ {\rm and}\ \ {\bf r}_{\scr 2}= {\bf R}-\frac{A_{\scr 1}}{A_{\scr P}}\ {\bf r},
\label{r1r2-Rr}
\end{equation}
where $A_{\scr 1}$ and $A_{\scr 2}$ stand for the mass numbers of the fragments, with $A_{\scr 1}+A_{\scr 2}=A_{\scr P}$. 

The total interaction between the projectile and the target is given by
\begin{equation}
V({\bf R},{\bf r})=V_{c_{\scr 1}-{\scr T}}(r_{\scr 1})+V_{c_{\scr 2}-{\scr T}}(r_{\scr 2}).
\label{eq:Vpt}
\end{equation}
The above potentials, which contain Coulomb and nuclear terms, are written as
\begin{equation}
V_{c_{\scr i}-{\scr T}}(r_{\scr i}) = V^{\scr (C)}_{c_{\scr i}-{\scr T}}(r_{\scr i}) + V^{\scr (N)}_{c_{\scr i}-{\scr T}}(r_{\scr i}),\ \ \ \ {\rm where}\ \ 
i=1,2.
\label{Vci-T_CN}
\end{equation}

For the Coulomb potentials, we adopt the usual approximation
\begin{eqnarray}
V^{\scr (C)}_{c_{\scr i}-{\scr T}}(r_{\scr i}) &=& \frac{Z_i\,Z_{\scr T}\, e^2}{r_i},\qquad\qquad\ \ \  r_i \ge R_{\scr C}, \nonumber \\
                                                                         &=&  \frac{Z_i\,Z_{\scr T}\, e^2}{2\,R_{\scr C}}\ 
                                                                         \left[ 3- \frac{r_{\scr i}^{\scr 2}}{R^{\scr 2}_{\scr C}} \right],\ \ \  r_i < R_{\scr C}.
\label{eq:uc}
\end{eqnarray}
Above, 
\begin{equation}
R_{\scr C} = r_{\scr 0C}\ \left( A_{\scr i}^{\scr 1/3}+ A_{\scr T}^{\scr 1/3} \right)
\label{R0C}
\end{equation}
is the Coulomb radius, where $r_{\scr 0C}$ is a parameter of the order of 1 fm. In our calculations we used the value, $r_{\scr 0C} = 1.06$ fm.

For the nuclear interactions $V^{\scr (N)}_{c_{\scr i}-{\scr T}}(r_{\scr i})$, different potential models are frequently used in heavy ion 
scattering~\cite{CaH13}. Among them, the Aky\"uz-Winther potential~\cite{BrW91,AkW81} is  a particularly convenient choice. This potential, 
based on the folding model, is usually approximated by a Woods-Saxon(WS) function of the form
\begin{equation}
V^{\scr (N)}_{c_{\scr i}-{\scr T}}(r_{\scr i}) = \frac{V_0}{1+\exp\left[ \left(r_i- R_{\scr 0N}\right)/a_{\scr N}\right]},
\label{WS}
\end{equation}
with $R_{\scr 0N} = r_{\scr 0N}\ \left( A_{\scr i}^{\scr 1/3}+ A_{\scr T}^{\scr 1/3} \right)$, and the parameters $V_0,r_{\scr 0N}$ and $a_{\scr N}$ 
are given by analytical expressions in terms of $A_{\scr i}$ and $A_{\scr T}$. \\

The total potential determines the classical trajectory of the projectile-target motion. It is also responsible for the couplings among the intrinsic states 
of projectile, leading to inelastic scattering and breakup reactions. For practical purposes, we split the potential of Eq.~(\ref{eq:Vpt}) into two parts, 
denoted by $V_{\rm opt}(R)$ and $\mathcal{V}({\bf R},{\bf r})$. The former depends exclusively on ${\bf R}$, so that it is diagonal in channel
space. It plays the role of the optical potential in nuclear reaction theory. This potential is used to determine the classical trajectory, ${\bf R}(t)$. The second
part, $\mathcal{V}({\bf R},{\bf r})$, is responsible for transitions to excited states of the projectile. The dependence of this potential on ${\bf R}$ can be 
transformed into dependence on $t$, by using the trajectory. That is, 
\begin{equation}\mathcal{V}({\bf R},{\bf r})\ \rightarrow \  \mathcal{V}(t,{\bf r})  \equiv  \mathcal{V}({\bf R}(t),{\bf r}).
\label{R to i}
\end{equation}

\smallskip

The separation of the total potential into an optical potential and a coupling interaction is somewhat arbitrary. However, once the former is chosen, the latter
is determined by the equation
\begin{equation}
\mathcal{V}({\bf R},{\bf r}) = V({\bf R},{\bf r}) - V_{\rm opt}(R).
\label{split}
\end{equation}
It is convenient to split $V_{\rm opt}(R)$ and $\mathcal{V}({\bf R},{\bf r})$ into the contributions from each fragment of the projectile, namely \\
\begin{eqnarray}
V_{\rm opt}(R) &=&  V_{\rm opt}^{\scr (1)}(R) +  V_{\rm opt}^{\scr (2)}(R)\\
\mathcal{V}({\bf R},{\bf r})  & &=   \mathcal{V}^{\scr (1)}({\bf R},{\bf r}) +  \mathcal{V}^{\scr (2)}({\bf R},{\bf r})\nonumber\\
\end{eqnarray}

We consider two choices of the optical potential, as described below.

\begin{itemize}

\item The two clusters are located at the center of the projectile (${\bf r}=0$). Then (see Eq.~(\ref{r1r2-Rr})),
${\bf r}_{\scr 1} = {\bf r}_{\scr 2} = {\bf R}$, and one gets
\begin{eqnarray}
V^{\scr (1)}_{\rm opt}(R) &=& V_{c_{\scr 1}-{\scr T}}(R) \label{option-old1}\\  
V^{\scr (2)}_{\rm opt}(R) &=& V_{c_{\scr 2}-{\scr T}}(R).
\label{option-old2}
\end{eqnarray}
This was the option adopted in Refs.~\cite{MCD08,MCD14}.
\item
The optical potential is given by the expectation value of the total potential, taken for the ground state of the projectile, $\Phi_0({\bf r})$. That is,
\begin{eqnarray}
V^{\scr (1)}_{\rm opt}(R) &=& \int d^3{\bf r}\   V_{c_{\scr 1}-{\scr T}} \left( r_{\scr 1}\right) \ \big| \Phi_0({\bf r}) \big|^2 
\label{option-new-1}\\
V^{\scr (2)}_{\rm opt}(R)  & =& \int d^3{\bf r}\  V_{c_{\scr 2}-{\scr T}} \left(r_{\scr 2}\right)
\Big]\ 
\big| \Phi_0({\bf r}) \big|^2.
\label{option-new-2}
\end{eqnarray}
where $r_{\scr 1}\equiv \left| {\bf r}_{\scr 1}\right|$ and $r_{\scr 2}\equiv \left| {\bf r}_{\scr 2}\right|$ are related to ${\bf R}$ and 
${\bf r}$ through Eq.~(\ref{r1r2-Rr}). This option is more appropriate for halo nuclei, and it is the one considered in the present work.

\end{itemize}

In some situations, the collision is influenced by intrinsic states of the target or by states of the projectile that have been left out of the
semiclassical coupled channel equations. These states remove flux from the incident current and this affects elastic/inelastic scattering 
and breakup. In this case, the sum of the occupation probabilities of the states included in the calculation should be less than one. 
This, effect may be simulated by the inclusion of an imaginary part in the optical potential.  This situation is not considered in the present
work.

\subsection{The semiclassical approach}

In the semiclassical method, the projectile-target motion is treated as a classical mechanics problem: the collision of two point-particles 
(the projectile and the target) interacting through the potential $V_{\rm opt}(R)$. In a collision with total energy $E$ and impact parameter 
$b$, the classical trajectory, ${\bf R}_b(t)$, is determined by solving classical equations of motion.\\

On the other hand, the intrinsic dynamics of the projectile is handled by full quantum mechanics, with the time-dependent Hamiltonian
\begin{equation}
h = h_0({\bf r}) + \mathcal{V}\left(b\,;{\bf r},t\right), 
\label{h}
\end{equation}
where $\mathcal{V}\left(b\,;{\bf r},t\right) \equiv \mathcal{V}\left({\bf R}_b{\small (t)},{\bf r}\right)$ is the sum of the interactions of the two clusters with the target, namely
\begin{equation}
\mathcal{V}\left({\bf R_{b}{\small (t)}},{\bf r}\right) = \mathcal{V}^{\scr (1)}\left(b\,; r_1(t) \right) + \mathcal{V}^{\scr (2)}\left(b\,; r_2(t) \right).
\label{Veq1}
\end{equation}
Above, 
\begin{equation}
r_{\scr 1} (t)= \sqrt{R_b^2(t)+\left(A_{\scr 2}/A_{\scr P}\right)^2\, r^{\scr 2}+2 \left(A_{\scr 2}/A_{\scr P}\right)\ {\bf R}_b(t)\cdot {\bf r}} 
\label{r_1(t)}
\end{equation}
and
\begin{equation}
r_{\scr 2} (t)= \sqrt{R_b^2(t)+\left(A_{\scr 1}/A_{\scr P}\right)^2\, r^{\scr 2}-2 \left(A_{\scr 1}/A_{\scr P}\right)\ {\bf R}_b(t)\cdot {\bf r}}
\label{r_2(t)}
\end{equation}
are the distances between the target and the clusters.

\subsubsection{The eigenstates of $h_0$}

The intrinsic Hamiltonian of the projectile has bound and unbound eigenstates. The unbound states, which are associated with the breakup channel, have 
negligible influence in collisions of tightly bound nuclei at near-barrier energies. However, the situation may be very different in collisions of
weakly bound nuclei. In this case, the breakup process affects strongly the reaction dynamics. Then, the influence of the continuum of the weakly
bound projectile has to be taken into account. \\

The intrinsic Hamiltonian is
\begin{equation}
h_0 = -\frac{\hbar^2}{2\mu_{\scr 12}}\ {\bf \nabla^2_r} + V_{c_{\scr 1}-c_{\scr 2}}(r) ,
\label{h0}
\end{equation}
where $\mu_{\scr 12}$  is the reduced mass of the two-cluster system and $V_{c_{\scr 1}-c_{\scr 2}}(r)$ is the
cluster-cluster interaction. This interaction can be written as
\begin{equation}
V_{c_{\scr 1}-c_{\scr 2}}(r) = V^{\scr (C)}_{c_{\scr 1}-c_{\scr 2}}(r) + V^{\scr (N)}_{c_{\scr 1}-c_{\scr 2}}(r),
\label{V12-a}
\end{equation}
where $V^{\scr (C)}$ and $V^{\scr (N)}$ are respectively the Coulomb and the nuclear terms. Of course, the former vanishes if one of the clusters 
is uncharged. Otherwise, the Coulomb potential is given by Eqs.~(\ref{eq:uc}) and (\ref{R0C}), with the replacements:
\[
Z_{\sc i}\,Z_{\scr T} \rightarrow Z_{\sc 1}\,Z_{\scr 2}; \ \ r_{\sc i} \rightarrow r; \ \  
R_{\scr C} \rightarrow R_{\scr 12} = r_{\scr 12}\, A_{\scr P}^{\scr 1/3} .
\]
In our calculations, we use $r_{\scr 12} = r_{\scr 0C} = 1.06$ fm.

\medskip

The nuclear interaction has the general form
\begin{equation}
V^{\scr (N)}_{c_{\scr 1}-c_{\scr 2}}(r) = - V_{\scr 12}\ f(r) + A_{\rm so}\  \left[ {\boldsymbol l}\cdot {\boldsymbol j}\right] \  \frac{1}{r}\  \frac{d f(r)}{dr} ,
\label{V12-b}
\end{equation}
where $f(r)$ is the Woods-Saxon function,
\begin{equation}
f(r) = \frac{1}{1+\exp[(r-R_{\scr 12})/a_{\scr 12}]},
\label{fr}
\end{equation}
In what follows, we assume that one of the clusters has spin zero and the other spin ${\boldsymbol j}$. In Eq.~(\ref{V12-b}), ${\boldsymbol l}$ is the 
orbital angular momentum of the $c_{\scr 1}-c_{\scr 2}$ relative motion and ${\boldsymbol J} = {\boldsymbol l} +{\boldsymbol j}$ is the total angular 
momentum of the projectile.  These angular momenta are measured in $\hbar$ units. In this potential,  $V_{\scr 12}$ 
is the depth of the volumetric term and $A_{\rm so}$ is the strength parameter of the spin-orbit term. Since $A_{\rm so}$ does not have the 
dimension of energy (it is an energy multiplied by a length squared), we do not use the notation $V_{\rm so}$.
For future use, to write Eq.~(\ref{V12-b}) as,
\begin{multline}
V^{\scr (N)}_{c_{\scr 1}-c_{\scr 2}}(r) =  - V_{\scr 12}\ f(r) + A_{\rm so}\  \left[ 
       \frac{1}{2}\,\left({\boldsymbol J}^2 - {\boldsymbol l}^2 - {\boldsymbol j}^2     \ \right) \right]\\
       \times\  \frac{1}{r}\  \ \frac{d f(r)}{dr}.
\label{V12-c}
\end{multline}

\bigskip\medskip
\noindent {\it i})\ {\it Bound states}
\bigskip

The bound eigenstates of the projectile satisfy the eigenvalue equation,
\begin{equation}
h_0 \ \Phi_\alpha ({\bf r}) = \varepsilon_\alpha  \ \Phi_\alpha ({\bf r}) ,
\label{eigen}
\end{equation}
where $\alpha$ stands for the set of quantum numbers required to specify the wave function. They satisfy the orthogonality condition,
\begin{equation}
\left\langle \Phi_\alpha \vert \Phi_{\alpha^\prime} \right\rangle = \delta_{\alpha , \alpha^\prime},
\label{ortho1}
\end{equation}
where $\delta_{\alpha , \alpha^\prime}$ is the Kronecker delta function.\\

Since one of the clusters has spin $j$ and the other has spin zero, the states are labeled by the set $\{\varepsilon_\alpha, j_\alpha,l_\alpha,
J_\alpha,M_\alpha\}$, where $j_\alpha$, $l_\alpha$ and $J_\alpha$ are respectively 
the principal quantum numbers of the spin, the orbital and the total angular momenta, whereas $M$ is the component of ${\bf J}$ along the z-axis.
Our calculations will neglect excitations of the cores. Therefore, the spin quantum number $j$ has a fixed value, independently of the state $\alpha$.
For this reason, we will omit it in the labels of the wave funcions. The orthogonal relation of Eq.~(\ref{ortho1}) then reads
\begin{eqnarray}
\left\langle \Phi_\alpha \vert \Phi_{\alpha^\prime} \right\rangle & \equiv&
\left\langle \Phi_{\varepsilon_\alpha l_\alpha J_\alpha M_\alpha}\vert \Phi_{\varepsilon_{\alpha^\prime} l_{\alpha^\prime} 
J_{\alpha^\prime} M_{\alpha^\prime}} \right\rangle  \nonumber \\
                & = & \delta_{\varepsilon_\alpha , \varepsilon_{\alpha^\prime}}\ \delta_{l_\alpha , l_{\alpha^\prime}}\ 
\delta_{J_\alpha , J_{\alpha^\prime}}\ \delta_{M_\alpha , M_{\alpha^\prime}}.
\label{ortho2}
\end{eqnarray}

\bigskip

As usual, we write the wave functions as products of a function of the modulus of ${\bf r}$ and another of its angular coordinates, represented by
$\hat{\bf r}$, in the form
\begin{equation}
\Phi_{\varepsilon_\alpha l_\alpha J_\alpha M_\alpha}({\bf r}) = \frac{\varphi_{\sc \varepsilon_\alpha l_\alpha J_\alpha}(r)}{r}\ \ \mathcal{Y}_{\sc l_\alpha J_\alpha M_\alpha}(\hat{\bf r}) ,
\label{RY}
\end{equation}
Since for each bound state the label $\alpha$ specifies the quantum numbers $\varepsilon_\alpha, l_\alpha, J_\alpha$ and $M_\alpha$, 
we can use the compact notation,
\begin{equation}
\Phi_{\alpha}({\bf r}) = \frac{\varphi_{\sc \alpha}(r)}{r}\ \ \mathcal{Y}_{\sc \alpha}(\hat{\bf r}) .
\label{RY-a}
\end{equation}
We remark that the situation is quite different for unbound states, where the energy quantum number, $\varepsilon$,  
is continuous. In this case there is an infinite number of degenerate states with energy $\varepsilon$, differing by the angular momentum quantum 
numbers $l,J$ and $M$.\\

The orientation-dependent part of the wave function is given by
\begin{equation}
\mathcal{Y}_{\alpha }(\hat{\bf r}) = \sum_{m\mu} \left\langle l_\alpha\, j\, m\,  \mu\vert J_\alpha M_\alpha 
\right\rangle \ 
Y_{l_\alpha\, m}(\hat{\bf r}) \, \left| j\mu\right\rangle,
\label{YlJM}
 \end{equation}
where $Y_{l_\alpha m}(\hat{\bf r})$ are spherical harmonics, $\left| j\mu\right\rangle$ are normalized spin states, and 
$\left\langle l_\alpha mj\mu\vert J_\alpha M_\alpha \right\rangle$ are Clebsh-Gordan coefficients. The functions of Eq.~(\ref{YlJM}) satisfy the 
orthonormality relation,
\begin{equation}
\left\langle \mathcal{Y}_{\alpha}\big| \mathcal{Y}_{\alpha^\prime}\right\rangle =
\delta_{ l_\alpha, l_{\alpha^\prime}}\ \delta_{ J_\alpha, J_{\alpha^\prime}} \ \delta_{ M_\alpha, M_{\alpha^\prime}}.
\label{orgho-Y}
 \end{equation}

\bigskip

The radial wave functions in Eq.~(\ref{RY-a}) satisfy the equation,
\begin{multline}
-\frac{\hbar^2}{2\mu_{\scr 12}}\ \left[ \frac{d^2}{dr^2} + \frac{l_\alpha \left( l_\alpha +1 \right)}{r^2} \right]
\varphi_{\alpha}(r)  \\
+\ V_{l_\alpha J_\alpha}(r)  \, \varphi_{\alpha}(r) = \varepsilon_\alpha\ 
\varphi_{\alpha}(r),
\label{bound-radeq}
\end{multline}
where $V_{l_\alpha J_\alpha}(r)$ is the potential of Eq.~(\ref{V12-a}) with the nuclear term of Eq.~(\ref{V12-c}), replacing the operators
${\boldsymbol l}^2$, ${\boldsymbol J}^2$ and ${\boldsymbol j}^2$ by their eigenvalues, $J_\alpha (J_\alpha+1),\, l_\alpha(l_\alpha+1)$ and
$j(j+1)$. \\

Normalizing the radial wave functions by the condition
\begin{equation}
\int_0^\infty\, dr\ \varphi^*_{\alpha}(r) \  \varphi_{\alpha^\prime}(r)
= \delta_{\varepsilon_\alpha,\varepsilon_{\alpha^\prime}},
\label{bound-normrad}
\end{equation}
one generates a set of orthonormal eigenstates of $h_0$, satisfying Eq.~(\ref{ortho2}).\\


\bigskip
\noindent {\it ii})\ {\it Unbound states}
\bigskip


The intrinsic Hamiltonian has also positive energy eigenstates, which correspond to  the elastic scattering of the two clusters. The scattering of 
an incident wave with wave vector ${\bf k}$ is an energy eigenstate with the continuous eigenvalue
\begin{equation}
\varepsilon = \hbar^2 k^2/2\mu_{\scr 12}.
\label{rel-ek}
\end{equation}

\medskip

In quantum scattering theory, the scattering wave function is expanded in partial-waves as
\begin{equation}
\Phi^{\scr (+)}_{\bf k}({\bf r}) = \sum_{\sc lJM} \Phi _{\sc \varepsilon lJM}({\bf r}),
\end{equation}
where~\cite{CaH13,Joa83}
\begin{equation}
\Phi _{\sc \varepsilon lJM}({\bf r}) = N\ \frac{u_{\sc k lJ}(r)}{kr}\ \mathcal{Y}_{\sc lJM}(\hat{\bf r}) .
\label{psi_klKM}
\end{equation}
Above, $N$ is a constant that ensures normalization of the wave function. \\

The radial wave function of Eq.~(\ref{psi_klKM}), $u_{\sc k lJ}(r)$, satisfies Eq.~(\ref{bound-radeq}) for the energy
of Eq.~(\ref{rel-ek}).
If both clusters are charged, the radial wave function is normalized as to have the asymptotic (outside the range of the short-range interaction) boundary condition
\begin{equation}
u_{\sc k lJ}(r) \ \longrightarrow\ \frac{i}{2} \ \Big[ H^{\scr (-)} (\eta ,kr) - \bar{S}_{\sc lJ} \,H^{\scr (+)} (\eta ,kr) \Big],
\label{uass-usual}
\end{equation}
and these wave functions satisfy the orthonormality relations,
\begin{equation}
\int dr\ u^*_{\sc k lJ}(r) \ \ u_{\sc k^\prime lJ}(r) = \frac{\pi}{2}\ \delta \left( k-k^\prime \right).
\label{ortho-u}
\end{equation}
In Eq.~(\ref{uass-usual}), $\bar{S}_{\sc lJ} $ is the nuclear S-matrix and $H^{\scr (-)} (\eta ,kr)$ ($H^{\scr (+)} (\eta ,kr)$) is the Coulomb wave 
function with ingoing (outgoing) wave boundary condition. The nuclear S-matrix is related to the nuclear phase-shift, $\delta_{\sc lJ}$, as
\begin{equation}
\bar{S}_{\sc lJ} = e^{2i\,\delta_{\sc lJ}}.
\end{equation}

\medskip

If one of the clusters is uncharged, the Coulomb term of the potential vanishes. The Coulomb wave functions in Eq.~(\ref{uass-usual}) are 
then replaced by the corresponding Riccati-Haenkel wave functions \cite{CaH13,Joa83}. 
If the cluster-cluster interaction is real, a proper choice of the normalization makes $u_{\sc k lJ}(r)$ real for all $r$. This can be 
achieved multiplying the radial wave function with the asymptotic behavior of Eq.~(\ref{uass-usual}) by the factor $\exp\left( -i\delta_l \right)$.\\

Owing to the continuous nature of the energy quantum number, the orthonormality  relations of Eq.~(\ref{ortho2}) must be modified. 
They become,
\begin{equation}
\left\langle \Phi_{\varepsilon  l  J  M} \vert \Phi_{\varepsilon^\prime l^\prime J^\prime M ^\prime} \right\rangle 
                 =  \delta \left( \varepsilon - \varepsilon^\prime \right)\ \delta_{l , l ^\prime} \ 
\delta_{J , J ^\prime} \ \delta_{M , M ^\prime} .
\label{ortho3}
\end{equation}

\bigskip

It is convenient to write the angular momentum projected unbound wave function similarly to Eq.~(\ref{RY}), namely,
\begin{equation}
\Phi_{\varepsilon  l  J  M} ({\bf r}) = \frac{z_{k  l  J}(r)}{r}\ \mathcal{Y}_{l  J  M}\left(\hat{\bf r}  \right),
\label{RY-1}
\end{equation}
with
\begin{equation}
z_{k  l  J}(r) = \left( \frac{2\mu_{\scr 12} }{\pi\hbar^2 k}\right)^{ 1/2}\ e^{-i\delta_{\sc lJ}}\ u_{k lJ}(k,r).
 \label{Rfunction}
\end{equation}
With the above normalization, the radial wave functions satisfy the orthonormality relations
\begin{equation}
\int dr\ z^*_{k  l  J}(r) \ z_{k^\prime  l  J}(r)= \delta\left(\varepsilon - \varepsilon^\prime \right),
 \label{Rfunction-norm}
\end{equation}
where $\varepsilon = \hbar^2 k^2/\mu_{\scr 12}$ and $\varepsilon^\prime = \hbar^2 k^{\prime 2}/\mu_{\scr 12}$. This guarantees that
Eq.~(\ref{ortho3}) is satisfied.

\subsubsection{The time-dependent coupled-channel equations}

In our semiclassical calculations we adopt a time scale such that the collision partners are at closest approach at $t=0$. In this way, 
the interaction potential goes to zero as $t\rightarrow \pm\infty$. Thus, the Hamiltonian of the projectile in these limits reduces to 
$h_0$. We assume that the projectile is initially ($t\rightarrow -\infty$) in its ground state, $\Phi_{0}
({\bf r})$. As the collision develops, the projectile interacts with the target and its wave function evolves according to the equation, 
\begin{equation}
\Big[ h_0 + \mathcal{V}(b;{\bf r},t) \Big] \ \Psi({\bf r},t) = i\hbar\  \frac{\partial }{\partial t} \Psi({\bf r},t).
\label{t-dep_Sch}
\end{equation}
To solve this equation we expand the time-dependent wave function in the eigenstates of $h_0$, as
\begin{multline}
\Psi({\bf r},t) = \sum_\alpha a_{\alpha} (b;t)\ \ e^{-i\varepsilon_\alpha\,t /\hbar}\ \ \Phi_{\alpha}({\bf r})\\
+\ \sum_{lJM} \int d\varepsilon\  a_{\varepsilon lJM }(b;t)\ \ e^{-i\varepsilon\,t /\hbar}\ \ \Phi_{\varepsilon\, l  J M}({\bf r}) .
\label{expansion}
\end{multline}
In collisions of tightly bound nuclei at near-barrier energies, the breakup channel can be neglected. The above expansion then 
reduces to its first line. Besides, the expansion can be truncated after a finite number of terms, say $N_{\scr B}$. Next, this expansion is 
inserted in Eq.~(\ref{t-dep_Sch}) and the resulting equation is multiplied by each $\left\langle \Phi_{\alpha} \right|$. In this way, one gets 
a set of $N_{\scr B}$ coupled differential equations for the amplitudes  $a_{\alpha} (b;t)$.
This procedure cannot be followed in collisions of weakly bound nuclei. In this case the breakup channel plays a major role in 
the collision dynamics, so that continuum states must be kept. In this way, the expansion is infinite, even truncating the energy 
at some limiting value $\varepsilon_{\rm max}$. \\

\bigskip
\noindent {\it i})\ {\it Continuum discretization: energy bins}
\bigskip

A reasonable way to deal with collisions of weakly bound nuclei is to use the CDCC approximation. It consists in approximating the 
radial wave functions in the continuum by a finite set of radial wave packets with constant (in time) shape. They are written as
\begin{equation}
\phi_{n l J}(r,t) = e^{-i\bar{\varepsilon}_n t/\hbar}\ \varphi_{n l J}(r) ,
\label{wpack-1}
\end{equation}
where the energy wave packets are
\begin{equation}
\varphi_{n l J}(r) = \int\ 
d\varepsilon\ \Gamma_n\left(\varepsilon\right)\ z_{\sc k l J}(r).
\label{wpack-22}
\end{equation}
Above, $k = \sqrt{2\mu_{\scr 12}\,\varepsilon}/\hbar$ and $\Gamma_n\left(\varepsilon\right)$ is a real function peaked around the energies $\varepsilon_n$, satisfying the orthonormality 
relations
\begin{equation}
\int_0^\infty d\varepsilon\ \Gamma_n\left(\varepsilon\right)\ \Gamma_{n^\prime} \left(\varepsilon\right) = \delta_{n,n^\prime}.
\label{ortho-Gamma}
\end{equation}
The time-dependent phase factors in Eq.~(\ref{wpack-1}) are given in terms of the energy expectation values
\begin{equation}
\bar{\varepsilon}_n = \int_0^\infty d\varepsilon\ \varepsilon\ \left| \Gamma_n\left(\varepsilon\right) \right|^2 .
\label{ebeta}
\end{equation}

\bigskip

With the wave packets of Eq.~(\ref{wpack-1}), we construct the full wave functions,
\begin{equation}
\Phi_{n l J M}({\bf r}) = \frac{\varphi_{n jJ}(r)}{r}\ \mathcal{Y}_{lJM}(\hat{\bf r}).
\label{CDCC basis}
\end{equation}
which are used as an approximate basis to describe the continuum space of the projectile. Owing to Eqs.~(\ref{Rfunction-norm}) and 
(\ref{ortho-Gamma}), the wave functions $\Phi_{n l J M}({\bf r})$ satisfy the orthonormality condition,
\begin{equation}
\left\langle \Phi_{n  l  J  M} \vert \Phi_{{n^\prime} l^\prime J^\prime M ^\prime} \right\rangle 
                 =  \delta_{n,n^\prime}\ \delta_{l , l ^\prime} \ 
\delta_{J , J ^\prime} \ \delta_{M , M ^\prime} .
\label{ortho4}
\end{equation}

\bigskip
\noindent {\it ii})\ {\it Bins in $k$-space}
\bigskip

In most applications it is more efficient to build wave packets in the wave number, $k$. For this purpose, one first modifies the normalization of
the radial wave functions as,
\begin{equation}
z_{k  l  J}(r) \rightarrow \tilde{z}_{k  l  J}(r) = \sqrt{\frac{2}{\pi}} \ \  e^{-i\delta_{\sc lJ}}\ u_{k lJ}(k,r).
 \label{Rfunction-1}
\end{equation}
With this change, the radial wave functions satisfy the orthonormality relations
\begin{equation}
\int dr\ \tilde{z}^*_{k  l  J}(r) \ \tilde{z}_{k^\prime  l  J}(r) = \delta\left( k-k^\prime \right).
 \label{k-ortho z}
\end{equation}
This result is trivially obtained, evaluating the above integral with the help of Eqs.~(\ref{Rfunction-1})  and (\ref{ortho-u}).

\bigskip

The bin wave functions have the general form of Eq.~(\ref{wpack-1}), as the bins in energy space. However, the wave packets are 
now given by the integrals in $k$,
\begin{equation}
\varphi_{n l J}(r) = \int\ 
dk\ \Gamma_n\left( k \right)\ \tilde{z}_{\sc k l J}(r).
\label{wpack-23}
\end{equation}
The generating functions, $\Gamma_n\left( k\right)$, are peaked around the wave numbers $k_n$, and they should satisfy
the relations
\begin{equation}
\int_0^\infty dk\ \Gamma_n\left(k\right)\ \Gamma_{n^\prime} \left(k\right) = \delta_{n,n^\prime}.
\label{ortho-Gamma-1}
\end{equation}
Using Eqs.~(\ref{wpack-22}) and (\ref{ortho-Gamma-1}), it is straightforward to prove that the bin states satisfy the orthonormality relations
\begin{equation}
\int dr\ \varphi^*_{n l J}(r) \ \varphi_{n^\prime l J}(r)  = \delta_{n,n^\prime} .
\label{ortho-bin}
\end{equation}
It follows immediately that the full states generated by the wave number bins (through Eq.~(\ref{CDCC basis})), satisfy
Eq.~(\ref{ortho4}).\\

The average energies of Eq.~(\ref{wpack-1}) are now given by
\begin{equation}
\bar{\varepsilon}_n = \int_0^\infty dk\ \left( \frac{\hbar^2 k^2}{2\mu_{\scr 12}} \right)\  \left| \Gamma_n\left(k\right) \right|^2 .
\label{kbeta}
\end{equation}
%


\subsection{The semiclassical CDCC equations}


With the discretization of the continuum, bound and unbound states can be treated in the same way. To stress this fact,
it is convenient to use labels $n$ (and $n^\prime$) to represent also bound states. In this case, $n$ stands for a state with
energy $\varepsilon_n$ and angular momentum quantum numbers $l_n,\, J_n$ and $M_n$. Since $n$ represent both bound
and bin states, it may take $N=N_{\scr B}+N_{\scr C}$ different values, running from 0 (the ground state of the projectile) to $N-1$. 
The states in the expansion are then written as
\begin{equation}
\Phi_n\left({\bf r}\right) = \frac{\varphi_n(r)}{r}\ \mathcal{Y}_n\left( \hat{\bf r}\right),
\label{expan-Phin}
\end{equation}
with $\mathcal{Y}_n\left( \hat{\bf r}\right)$ given by Eq.~(\ref{YlJM}) (with $\alpha$ replaced by $n$). In the case of bound states, 
the radial wave functions are the normalized solutions of the radial equation (Eq.~(\ref{bound-radeq}) with the replacement $\alpha\rightarrow n$), 
whereas for unbound states they are the bin wave functions of Eq.~(\ref{wpack-22}), in the case of bins in the $\varepsilon$-space, or Eq.~(\ref{wpack-23}), 
for discretization
in the $k$-space. \\

The expansion of the time-dependent wave function can be written in the compact form,
\begin{equation}
\Psi\left({\bf r},t \right) = \sum_{n^\prime=0}^{N-1}\,a_{n^\prime}(b,t)\ e^{-i\,\varepsilon_{n^\prime} t/\hbar}\ \Phi_{n^\prime}({\bf r}).
\label{exp-n}
\end{equation}
Inserting the above equation in Eq.~(\ref{t-dep_Sch}), and taking scalar product with each state $\left\langle \Phi_n \right|\, 
\exp\left( i\varepsilon_n\,t/\hbar\right)$,  one gets the set of coupled equations,
\begin{eqnarray}
i\hbar\ \dot{a}_0(b,t) &=& \sum_{n^\prime = 0}^{N-1}\ e^{i \left(\varepsilon_0-\varepsilon_{n^\prime}  \right) t/\hbar}\ 
\mathcal{V}_{ 0,n^\prime}(b,t)\ a_{n^\prime}(b,t) \nonumber \\
.................& &......................................................\nonumber\\
i\hbar\ \dot{a}_N(b,t) &=& \sum_{n^\prime = 0}^{N-1}\ e^{i \left(\varepsilon_N-\varepsilon_{n^\prime}  \right) t/\hbar}\ 
\mathcal{V}_{ N, n^\prime}(b,t)\ a_{n^\prime}(b,t) , \nonumber \\
& &
\label{CC}
\end{eqnarray}
where the coupling matrix-elements are 
\begin{equation}
\mathcal{V}_{ n, n^\prime}(b,t) = \big\langle \Phi_n \big|\, \mathcal{V} \big( {\bf r}, {\bf R}_b(t) \big)\, \big| \Phi_{n^\prime} \big\rangle.
\label{Vcoup-3}
\end{equation}
A detailed discussion of these matrix-elements is presented in the appendix.

\subsubsection{Detailed discussion of the bin wave packets}


For discretizations in the $\varepsilon$-space, the functions $\Gamma_\beta(\varepsilon)$ are concentrated in the vicinity of the expectation values of 
Eq.~(\ref{ebeta}). At energies far from $\varepsilon_\beta$ they are vanishingly small. The continuum is truncated at some maximal energy, 
$\varepsilon_{\rm max}$ and the continuum states with energies in the interval $\{ 0,\varepsilon_{\rm max} \}$ are approximated by a set of $N$ 
wave packets, with labels $n=1,..., N$. Frequently, the wave packets are generated by weight functions with the form
\begin{equation}
\Gamma_n\left(\varepsilon\right) = \frac{1}{\sqrt{\Delta_n}}\ \Big[ \Theta(\varepsilon - \varepsilon_{n_-}) - \Theta(\varepsilon -\varepsilon_{n_+})   \Big],
\end{equation}
where $\Theta(\varepsilon - \varepsilon_{n_-})$ and $\Theta(\varepsilon -\varepsilon_{n_-})$ are Heaviside step functions. Above, $\varepsilon_{n_+}$ 
and $\varepsilon_{n_+}$ are respectively the lower and the upper limits of the bin, and $\Delta_n = 
\varepsilon_{n_+}-\varepsilon_{n_-}$. It can be easily checked that $\varepsilon_{n_\pm} = \varepsilon_{n}\pm \Delta_n/2$.
The width of the $\Gamma_n(\varepsilon_n)$ functions, $\Delta_n$, is correlated with the extension of the wave packet of Eq.~(\ref{wpack-1}) in the coordinate space. 
For sharp $\Gamma_n(\varepsilon)$, like that with $\Delta_n = 25$ keV, the wave packets reach more than 1000 fm. In principle, the generating functions have  
have a constant width and their maxima are equally spaced. However, a sharper generating functions must be used in the vicinity of resonances. On the other hand, discretizations in the $k$-space with a constant width correspond to the generating functions with increasing energy widths. \\

Bertulani and Canto~\cite{BeC92} have shown that the wave packets converge more smoothly if one avoids the sharp edges of
the above step functions. They have built wave packets in momentum space based on the functions,
\begin{equation}
\chi_j (k) = N_j\ \left( \frac{k}{\Delta}\right)^{n_j^2}\ e^{-n_j\, k/\Delta}
\label{chi-set}
\end{equation}
where  $n_j = m \times j$, and $N_j$ is the normalization constant
\begin{equation}
N_j = \frac{1}{\sqrt{\Delta}}\ \left[ 
\frac{\left( 2n_j \right)^{2n_j^2+1}} {\left( 2n_j^2 \right) !} \right]^{1/2} .
\label{normal-chi}
\end{equation}
The above set of momentum states is characterized by two parameters: the integer $m$, which determines the spacing of two consecutive states, 
and the parameter $\Delta$, which gives their widths.  It can be easily checked that these functions are peaked at $k = n_j\,\Delta$, and that they 
have average momentum $\left< k \right> = \Delta\, \left( 2n_j^2+1 \right)/2 n_j$.\\

Eq.~(\ref{normal-chi}) guarantees that the states $\chi_j$ are normalized, although they are not orthogonal. However, these states
can be used to build an orthonormal set. This can be achieved by using the Gram-Schmidt method~\cite{ByF69}. The orthonormal 
set is given by
\begin{equation}
\Gamma_j (k) = \sum_{j^\prime} c_{j,j^\prime}\,\chi_{j^\prime}(k),
\label{Gamma-set}
\end{equation}
with
\begin{eqnarray}
c_{j,j^\prime} &=& -\left< \chi_{j^\prime}\big| \chi_{j} \right>,\ \ {\rm for\ } j^\prime<j \nonumber\\
                      &=&0,  \qquad\qquad\ \ \  {\rm for\ } j^\prime \ge j. 
\end{eqnarray}

\section{Applications}

We use the semiclassical method of the previous sections to evaluate CF and TF cross sections in collisions of $^{6,7}$Li projectiles on $^{197}$Au and $^{165}$Tb targets,
for which experimental data are available. \\

The first step in our calculations is to choose the parameters of the intrinsic Hamiltonian of the $^6$Li and $^7$Li nuclei, and to solve the Schr\"odinger equation for
each projectile, finding their bound and continuum eigenstates. For each nucleus, the parameters of the potential were determined by the condition of reproducing the 
experimental energies of the bound states and resonances. In the case of $^6$Li, treated as a $^4$He\,+\,$^2$H} system, the g.s. with $J^\pi = 1^+$ 
($l=0$ coupled to $s=1$) and binding energy $\varepsilon_0 = -1.47$ MeV is the only bound state. We considered also the resonances with $J^\pi = 3^+, 2^+$ and $1^+$ 
resulting from the coupling of $l=2$ with $s=1$. The $^7$Li nucleus, treated as a $^4$He\,+\,$^3$H system, has two bound states. The g.s., with $J^\pi = 3/2^+$ ($l=1$
coupled to $s=1/2$) with energy $\varepsilon_0 = -2.47$ MeV and an excited state with $J^\pi = 1/2^-$ with energy $\varepsilon_1 = -1.99$ MeV. In addition, we considered
the resonances with $J^\pi = 7/2^-$ and $5/2^-$, corresponding to couplings of $l=3$ with $s=1/2$. The parameters of $h_0$ for each nucleus were determined by the
condition of reproducing the energies of the bound states and resonances. Note that the parameters for the bound states and for the continuum are not necessarily the same. 
Since the details of this procedure and the resulting potential parameters are presented in Ref.~\cite{DTB03}, they are omitted here.\\

In the discretization of the continuum, we used eigenstates of $h_0$ with energies up to $\varepsilon_{\rm max} = 8$ MeV and angular momenta up to $4\hbar$, in the case 
of $^6$Li, and $5\hbar$, in the case of $^7$Li. The space spanned by these states was large enough to guarantee convergence in the CDCC calculations of Diaz-Torres, 
Thompson and Beck~\cite{DTB03}. The bins were then wave packets in this space, generated by orthonormal auxiliary functions in momentum representation, $\Gamma_j (k) $, 
given by Eq.~(\ref{Gamma-set}). Here and in the subsequent calculations of the cross sections, we adopt the width parameters $\Delta = 0.02984\ {\rm fm}^{-1}$, for $^6$Li, and 
$\Delta = 0.03384\ {\rm fm}^{-1}$, for $^7$Li, and take $m=3$. This means that the spacing between the maxima of $\chi_j (k)$ and of $\chi_{j\pm 1} (k)$ is three times the 
width $\Delta$. Using a constant $\Delta$ in $k$-space, the energy width of the bins and the energy spacing between consecutive bins grow linearly with $k$. \\

For the nuclear interactions between the fragments of the $^{6,7}$Li projectiles and the target (Eq.~(\ref{Vci-T_CN})), we use Aky\"uz-Winther potentials~\cite{BrW91,AkW81}, and total 
projectile-target interaction, $V({\bf R},{\bf r})$, is given by the sum of the two fragment-target potentials (see Eq.~(\ref{eq:Vpt})). The optical potential, which is used in the 
calculation of classical trajectories is given by the expectation value of $V({\bf R},{\bf r})$ with respect to the g.s. of the projectile. That is,
\begin{equation}
V_{\rm opt}(R) = \int d{\bf r}\   \Big[ V_{c_{\scr 1}-{\scr T}}(r_1) +V_{c_{\scr 2}-{\scr T}}(r_2) \Big]\ \Big| \Phi_0(r) \Big|^2.
\label{Voptical}
\end{equation}
The trajectory for each collision energy and impact parameter, $R_b(t)$, is an essential ingredient in the semiclassical theory. It transforms the ${\bf R}$-dependence
of the operator $V({\bf R},{\bf r})$ into time-dependence. The off-diagonal part of this operator, denoted by $\mathcal{V}({\bf R},{\bf r})$, is then expanded in multipoles up to
$\lambda =6$, and the matrix-elements appearing in the semiclassical coupled-channel equations are evaluated as explained in the appendix. These matrix-elements 
involve radial integrals over $r$, which are evaluated between $r=0$ and $r=r_{\rm max} = 1000$ fm.\\

The classical trajectories, $R_b(t)$, evolve independently
of the quantum mechanical amplitudes. That is, for each impact parameter and collision energy, the trajectory is determined by the classical equations of motion,
which disregard the intrinsic degrees of freedom of the system. In this way, energy conservation is violated. This was not a serious problem in early calculations of
inelastic scattering to rotational states~\cite{AlW75}, where the intrinsic excitations are negligible in comparison to the collision energy. Here, however, the situation 
is more complicated, since some bins have excitations comparable to the collision energy. Unfortunately, violation of energy conservation is an intrinsic drawback 
of the semiclassical approximation. In fact, it may even lead to the population of closed channels. To remedy this extreme situation, we adopt the following
procedure. At each collision energy and impact parameter, we solve the classical equations of motion, up to $t_f$. Next, we evaluate the radial kinetic 
energy of the relative motion at the end point, $K_{\scr R}$. Since the population of continuum states occur mostly in the neighbourhood of $t_f$,  $K_{\scr R}$ is the
energy available for transfer to intrinsic degrees of freedom.  We then reset  $\varepsilon_{\rm max} = K_{\scr R}$, if $K_{\scr R} < \varepsilon_{\rm max}$, and run the calculation of the amplitudes $a_n(t)$. 
In this way, we eliminate the excitation of closed channels.\\

The optical potential of Eq.~(\ref{Voptical}) is also used in the calculation of the tunnelling probabilities, that appears in the expression for the CF cross section. An important feature 
of this potential is that it takes into account the static effects of low breakup thresholds on complete fusion. The barrier of this potential tends to be lower than that of an AW 
potential between the whole projectile and the target, mainly in the case of weakly bound systems. This is illustrated in table \ref{Tab1}, which shows the barrier of the two potentials
for each system studied in this work. The barriers associated with the optical potentials of Eq.~(\ref{Voptical}), denoted by $V_{\scr B}^{\scr (2)}$, are shown in the fourth 
column whereas the barrier of the
projectile-target AW potentials, denoted by $V_{\scr B}^{\scr (1)}$, appear in the third column. The last column of the table, $\Delta V$, is the difference between the two. It gives the 
barrier lowering due to the weak binding of the projectile. As expected, the barrier lowering for collisions of $^6$Li (breakup threshold $B = 1.47$ MeV), is much more important 
than in the case of $^7$Li (breakup threshold $B = 2.47$ MeV).
\begin{table}
\caption{Potential barriers for the systems studied in the present paper. The contents of the columns are explained in the text.
}
\centering
\begin{tabular} [c] {ccccc}
\hline 

\ \ \ \ \ P\ \ \ \ \         &\ \ \ \ \   T\ \ \ \ \            &   \ \ \ \ \  $V_{\scr B}^{\scr (1)}$\ \ \ \ \      &  \ \ \ \ \  $V_{\scr B}^{\scr (2)}$\ \ \ \ \           &   \ \ \ \ \   $\Delta V$    \ \ \ \ \     \\
 \hline 
 $^6$Li            &     $^{159}$Tb        &           25.0                            &                                     23.7                                                            &         1.3                    \\                                                                 
%
 $^7$Li            &     $^{159}$Tb        &           24.7                            &                                     24.4                                                            &         0.3                    \\                                                                 
%
 $^6$Li            &     $^{197}$Au        &           29.3                            &                                     27.8                                                            &         1.5                    \\      
 %
 $^7$Li            &     $^{197}$Au        &           28.9                            &                                     28.6                                                            &         0.3                   \\              
    \hline 
\end{tabular}
\label{Tab1}
\end{table}

\bigskip

For practical purposes, one has to solve the system of of coupled equations for the amplitudes $a_n(t)$, Eq.~(\ref{CC}). 
This is done numerically, starting from some point $t=t_0$, where $R_{\rm in} \equiv R_b\left(t_0\right)$ is large enough
for the coupling matrix elements to be negligible.
We use $R_{\rm in} = 200$ fm. 
At this point the amplitudes  have their initial values, $a_n\left(t_0\right) =\delta_{n0}$. 
The integration runs until a final $t_f$, where the trajectory reaches the point of closest 
approach (if $E < V_{\scr B}$) or the strong absorption radius (if $E > V_{\scr B}$). 
The final values of the amplitudes are then used in the calculations of the cross sections. \\

The integral over impact parameter was transformed into a sum over integer values of the angular momentum 
associated with the projectile-target motion, $L$ (in $\hbar$ units). 
We used the classical relation $L = K b$, with $K=\sqrt{2E_{\rm cm}\mu_{\scr pt}}/\hbar$, with $\mu_{\scr pt}$ standing for 
the reduced mass of the projectile-target system and $E_{\rm cm}$ for the collision energy in the center of mass frame. 
This sum was truncated at  $L_{\rm max} = 60$. 
We remark that this angular momentum cut-off would be too small in calculations of breakup cross sections, due to the 
long range of Coulomb couplings. 
However, large angular momenta correspond to distant collisions, which do not contribute to fusion. 
Thus,  $L_{\rm max} = 60$ is more than enough for the calculations of the present work.\\


\subsection{Study of $^{6,7}$Li + $^{159}$Tb fusion}


%
\begin{figure}
\includegraphics[width=7.5 cm]{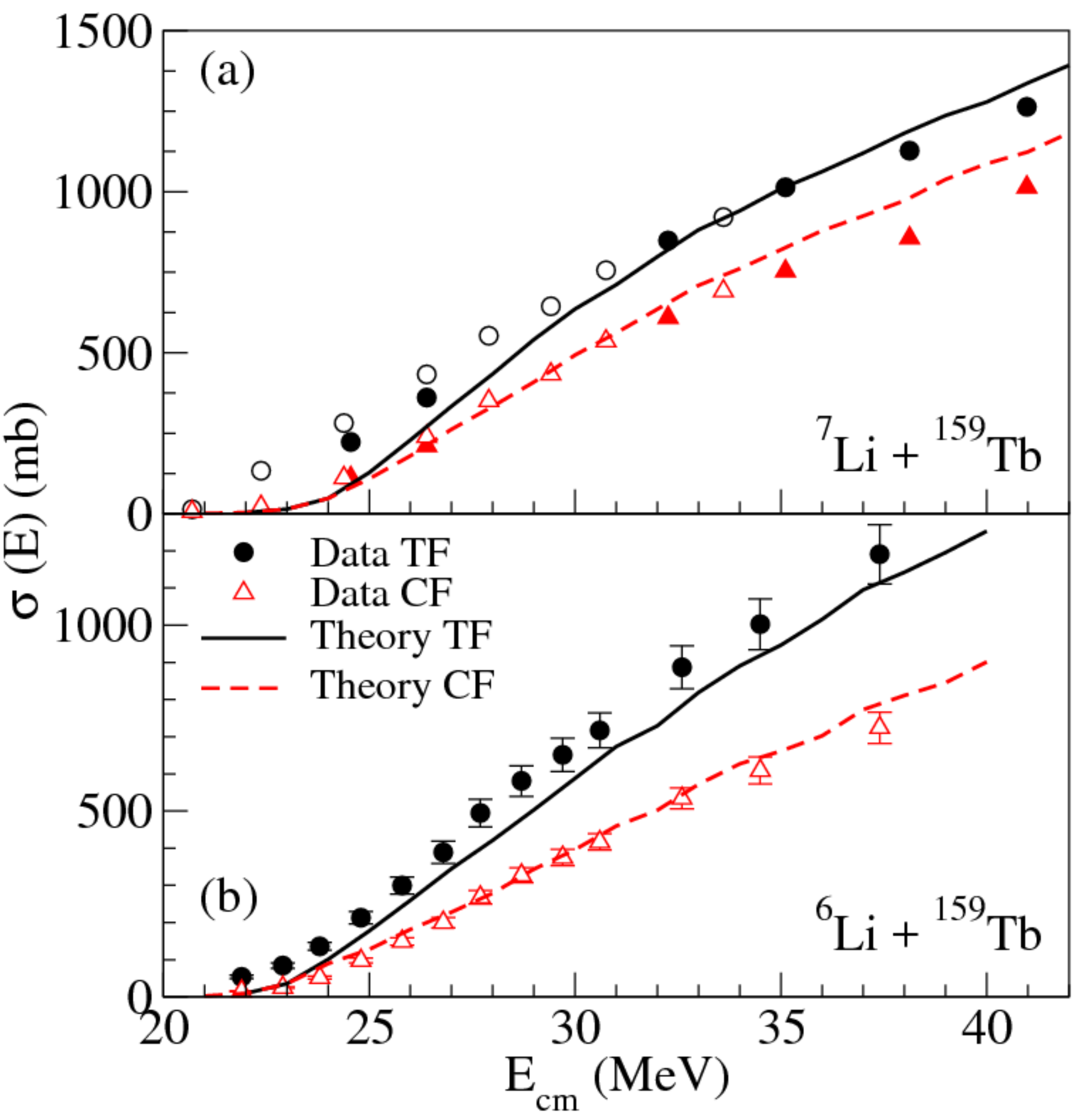}
\caption{(color on line) Semiclassical CF and TF cross sections for the $^{7}$Li + $^{159}$Tb (panel (a)) and 
$^{6}$Li + $^{159}$Tb (panel (b)) systems. 
For comparison, the experimental data of Mukherjee {\it et al.}~\cite{MSP06}  (full symbols) and Broda {\it et al.}~\cite{BIH75} (hollow symbols) 
are shown in panel (a), and those of Pradhan {\it et al.}~\cite{PMB11} are exhibited in panel (b). 
Triangles and circles denote respectively CF and TF data in both panels.}
\label{159Tb}
\end{figure}
CF and TF cross sections for the $^{7}$Li + $^{159}$Tb system are shown on panel (a) of Fig.~\ref{159Tb}. 
The red dashed line and the black solid line represent respectively the CF and the TF cross sections obtained with our semiclassical model. 
The figure shows also the experimental CF (triangles) and TF (circles) cross sections of Refs.~\cite{PMB11,MSP06}.
One concludes that the theoretical CF cross sections describes very well the experimental data, although they overestimate slightly the data 
at the highest energies considered. 
In the case of TF, the agreement between theory and experiment is worse, especially at the lower energies of the $^{7}$Li + $^{159}$Tb system. 
In this region, the theoretical cross section is appreciably lower than the data. 
Since the theoretical CF cross section agrees with the data, the difference must arise from ICF, which results from the breakup of the projectile. 
We find that this experimental result deserves further attention, since it is not obvious to us how the  $^{7}$Li projectile could give rise to a higher 
ICF cross section than the more weakly bound $^{6}$Li nucleus. 
The theoretical calculations predict a higher ICF cross section in the case of $^{6}$Li, as one would expect based on the break up mechanism. \\

\begin{figure}
\includegraphics[width=8 cm]{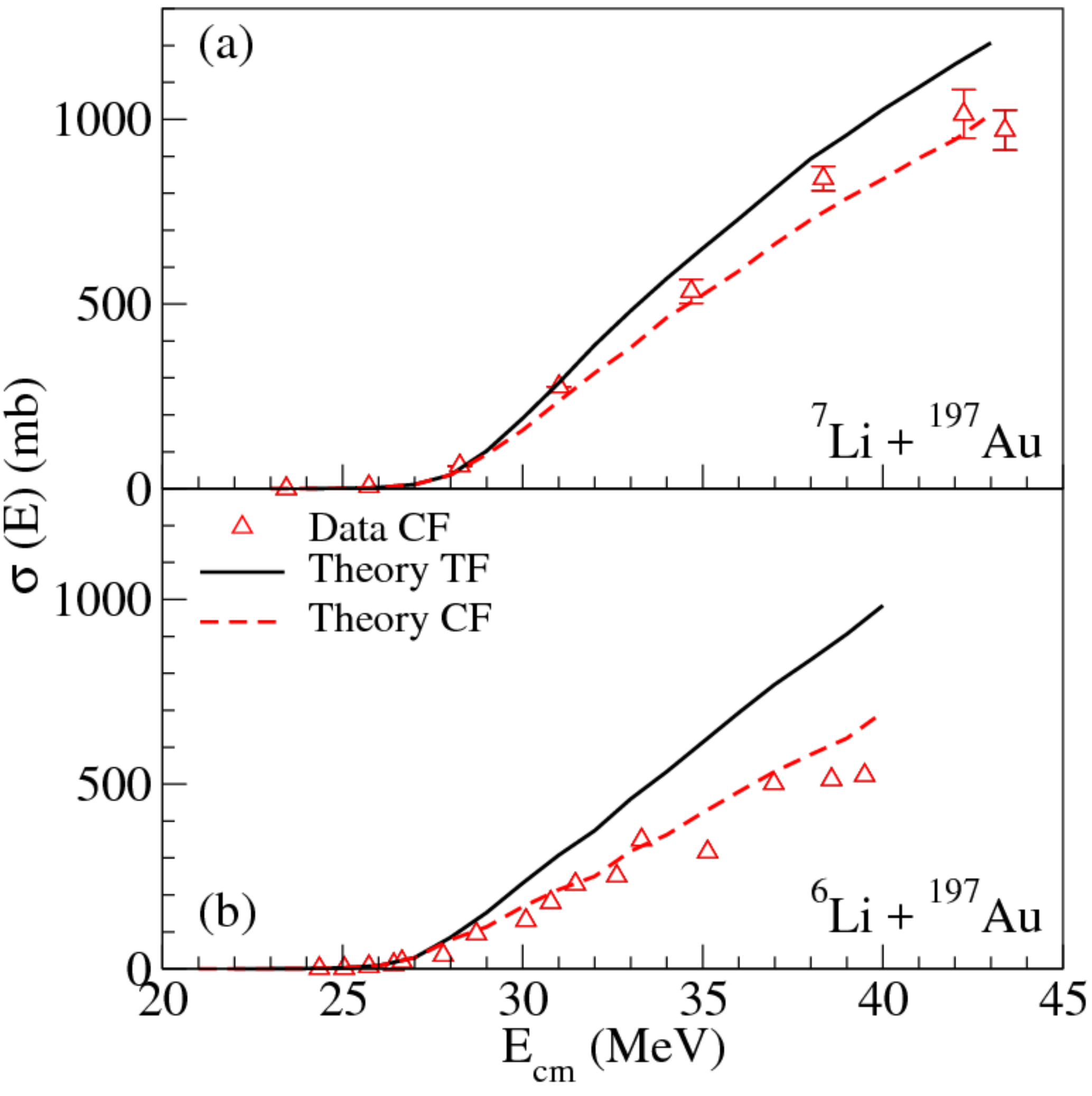}
\caption{(color on line) Similar to the previous figure but now the target is  $^{197}$Au. 
The CF data is from Palshetkar {\it et al.}~\cite{PTN14}. In this case, TF data is not available.
The notation is the same as in the previous figure.}
\label{197Au}
\end{figure}
A similar study for the $^{6}$Li + $^{159}$Tb system is presented on panel (b) of the same figure. 
Now the theoretical CF cross section reproduces the data extremely well. 
In the case of TF, the agreement is also good, although the theoretical cross section is slightly below the data. 
We remark that this potential takes into account the low binding energy of weakly bound projectiles, reducing 
the height of the Coulomb barrier. 
This is a significant effect in collisions of the $^6$Li projectile that has a very low breakup threshold. 
This is illustrated in table \ref{Tab1}.

\subsection{Study of $^{6,7}$Li + $^{197}$Au fusion}

Figure \ref{197Au} shows CF and TF cross sections for the $^{6,7}$Li + $^{197}$Au systems. Again, the red dashed lines and the black solid lines represent respectively CF and TF cross
sections calculated by the semiclassical model. The red triangles are the experimental CF cross sections of Palshetkar {\it et al.}~\cite{PTN14}. There are no experimental TF data for
these systems. The cross sections for $^{7}$Li + $^{197}$Au are shown on panel (a) whereas the ones for $^{6}$Li + $^{197}$Au are shown on panel (b). In both cases the agreement
between theory and experiment is good. 


\section{Conclusions}


We used an improved version of the semiclassical method to investigate complete and total fusion in collisions of weakly bound
nuclei. The total interaction between the collision partners is given by the sum of Aky\"uz-Winther potentials between the target
and each of the projectile's fragments. The projectile-target trajectory is then determined by classical equations of motion, with the
potential given by the expectation value of this interaction with respect to the ground state of the projectile. This potential is also
used to evaluate the tunnelling probabilities in the expression for the CF cross section. In this way, static barrier lowering effects 
associated with the low breakup threshold of the projectile are taken into account. \\

In the semiclassical method, the incident energy is treated as a constant of motion along the trajectory. On the other hand, the internal energy
of the projectile increases, as excited states are populated. Thus, the method does not conserve the total energy of the system. In this way, 
it is possible to excite even closed channels. To avoid such extreme situations we introduced a cut-off energy in the continuum, limiting it to the 
available kinetic energy of the relative motion. This procedure has led to reasonable results, although more sophisticated approaches are possible
and presently being considered. \\

We performed calculations of CF and TF cross sections for collisions of $^{6,7}$Li projectile with $^{159}$Tb and $^{197}$Au targets.
In each case, the theoretical cross sections were compared with the available data. The overall agreement between theory and experiment
is fairly good. 

\section*{Acknowledgments}
We thank Dr.\ A.\ Mukherjee for providing the data for the $^6$Li+$^{159}$Tb system.
Work supported in part by the Conselho Nacional de Desenvolvimento Cient\'\i fico e Tecnol\' ogico (CNPq), Coordena\c c\~ao de Aperfei\c coamento de Pessoal de N\'\i vel Superior (CAPES), Funda\c c\~ao Carlos Chagas Filho de Amparo \`a Pesquisa do Estado do Rio de Janeiro (FAPERJ), a BBP grant from the latter, the Programa de Desarrollo de las Ciencias B\' asicas (PEDECIBA) and the Agencia Nacional de Investigaci\'on e Innovaci\' on (ANII). 
This work has been done as a part of the project INCT-FNA, Proc. No.464898/2014-5.
We also thank the N\'ucleo Avan\c cado de Computa\c c\~ao de Alto Desempenho (NACAD), Instituto Alberto Luiz Coimbra de P\'os-Gradua\c c\~ao e Pesquisa em Engenharia (COPPE), Universidade Federal do Rio de Janeiro, for the use of the supercomputer Lobo Carneiro, where part of the calculations has been carried out.


\appendix

\section{Evaluation of matrix-elements for the semiclassical CC equations}

\bigskip


In this appendix we presented a detailed derivation of the angular momentum projected matrix-elements that appear in the semiclassical
CDCC equations. \\

The first step to evaluate the coupling matrix-elements is to carry out a multipole expansion of the interaction of Eq.~(\ref{Veq1}) as, 
\begin{equation}
\mathcal{V}({\bf R},{\bf r}) = \frac{1}{4\pi}\ \sum_{\lambda}\, \mathcal{V}^{(\lambda)}(R,r)\ \sum_\nu \, Y_{\lambda\, \nu}
(\hat{\bf R})\  Y^*_{\lambda\, \nu} ({\hat\bf r}).
\end{equation}
The matrix elements of Eq.~(\ref{Vcoup-3}) can then be evaluated using the above multipole expansion and the factorized wave functions of 
Eq.~(\ref{expan-Phin}). One gets,
\begin{multline}
\mathcal{V}_{n,n^\prime}(b;t) =\sum_{\lambda}\, I^{(\lambda)}_{n,n^\prime}(R)\  \sum_\nu \Big[  Y_{\lambda\, \nu} (\hat{\bf R})\\
\times  \int \Omega_{\hat{\bf r}}\ \mathcal{Y}^\dagger_n\left( \hat{\bf r} \right)\ 
Y^*_{\lambda\, \nu} (\hat{\bf r})\  \mathcal{Y}_{n^\prime}\left( \hat{\bf r} \right) \Big],
\label{mat-element}
\end{multline}
where $I^{(\lambda)}_{n,n^\prime}(R)$ is the radial integral,
\begin{equation}
I^{(\lambda)}_{n,n^\prime}(R) =
 \int dr\ \varphi^*_n(r)\ \
\frac{\mathcal{V}^{(\lambda)}(R,r)}{4\pi}\ \
\varphi_{n^\prime}(r).
\label{intrad}
\end{equation}
The factor given by the angular integral in Eq.~(\ref{mat-element}) is more complicated. It will be evaluated in the sub-sections below.

\subsection{A simpler situation: two spin zero clusters}

In this case, the angular part of the wave function reduces to the spherical harmonics, and Eq.~(\ref{mat-element})  can be written as,
\begin{equation}
\mathcal{V}_{n,n^\prime}({\bf R}) = \sum_\lambda I^{(\lambda)}_{n,n^\prime}(R)\ 
\sum_\nu\ Y_{\lambda\,\nu}( \hat{\bf R} ) \ \ \mathcal{B}^{(\lambda\nu)}_{l_n m_n,l_{n^\prime} m_{n^\prime} }\ .
\label{mat-el-2}
\end{equation}
Above, $\mathcal{B}^{(\lambda\nu)}_{l_n m_n,l_{n^\prime} m_{n^\prime} }$ is the geometric factor, 
\begin{equation}
\mathcal{B}^{(\lambda\nu)}_{l_n m_n,l_{n^\prime} m_{n^\prime}} = \int d\Omega_{\hat{\bf r}}\ Y^*_{l_n\,m_n}(\hat{\bf r})\ 
Y_{\lambda\,\nu}(\hat{\bf r}) \  
Y_{l_{n^\prime}\,m_{n_\prime}}(\hat{\bf r}).
\label{Blm-1}
\end{equation}
Since $Y^*_{l \,m}(\hat{\bf r}) = (-)^m\ Y_{l \,- m}(\hat{\bf r})$, the above expression can be put in the form,
\begin{multline}
\mathcal{B}^{(\lambda\nu)}_{l_n m_n,l_{n^\prime} m_{n^\prime}}   = (-)^{m_n} \int d\Omega_{\hat{\bf r}}\ 
 Y_{l_n\,-m_n}(\hat{\bf r})\\ 
 \times\ Y_{\lambda\,\nu}(\hat{\bf r}) \,   
Y_{l_{n^\prime}\,m_{n^\prime}}(\hat{\bf r}).
\label{Blm-2}
\end{multline}
To evaluate the above integral, we use the well known relation,
\begin{multline}
\int d\hat{\bf r}\ Y_{j_1\, m_1}(\hat{\bf r})\ Y_{j_2\, m_2}(\hat{\bf r})\ Y_{j_3\, m_3}(\hat{\bf r})= \\ \sqrt{ \frac{
\left(2j_1+1\right)\, \left(2j_2+1\right)\, \left(2j_3+1\right)}{4\pi} }\\   \\  \times\ \left(
\begin{array}
[c]{ccc}%
j_1\, & j_2 & j_3\\
0 & 0 & 0
\end{array}
\right) \
\left(
\begin{array}
[c]{ccc}%
j_1\, & j_2 & j_3\\
m_1 & m_2 & m_3
\end{array}
\right).
\end{multline}
We get,
\begin{multline}
\mathcal{B}^{(\lambda\nu)}_{l_n m_n,l_{n^\prime} m_{n^\prime}}    = (-)^{m_n}\ 
\sqrt{ \frac{
\left(2l_n+1\right)\, \left(2\lambda+1\right)\, \left(2l_{n^\prime}+1\right)}{4\pi}
}\\ 
\\
\times\
\left(
\begin{array}
[c]{ccc}%
l_n\, &\lambda &l_{n^\prime}\\
0 & 0 & 0
\end{array}
\right) \
\left(
\begin{array}
[c]{ccc}%
l_n\, &\lambda &l_{n^\prime}\\
-\,m_n & \nu & m_{n^\prime}
\end{array}
\right)\ \delta_{m_{n^\prime} -m_n + \nu}.
\label{Blm-3}
\end{multline}
The above result can also be expressed in terms of Clebsh-Gordan coefficients. Using the relation
relation
\begin{equation}
\left(
\begin{array}
[c]{ccc}%
j_1\, & j_2 & j_3\\
m_1 & m_2 & -m_3
\end{array}
\right) =\frac{ (-)^{j_1-j_2+m_3}}{\sqrt{2j_3+1}}\ \Big\langle j_1\, j_2\,m_1\, m_2\,\Big|\, j_3\,m_3\Big\rangle,
\end{equation}
one obtains,
\begin{multline}
\mathcal{B}^{(\lambda\nu)}_{l_n m_n,l_{n^\prime} m_{n^\prime}}    = (-)^{m_n-m_{n^\prime}}\ 
\sqrt{ \frac{\left(2l_n+1\right)\, \left(2\lambda+1\right)}{4\pi\, \left(2l_{n^\prime}+1\right)}}\\ 
\times \left\langle l_n\, \lambda\, 0\, 0   \vert l_{n^\prime} 0     \right\rangle\
\left\langle l_n\, \lambda\, m_n\,  \nu \vert l_{n^\prime} m_{n^\prime}    \right\rangle.
\label{Blm-4}
\end{multline}
%
\subsection{Matriz-elements when one cluster has spin ${\boldsymbol j}$}

In this case Eq.~(\ref{mat-el-2}) becomes,
\begin{equation}
\mathcal{V}_{n,n^\prime}({\bf R}) = \sum_\lambda I^{(\lambda)}_{n,n^\prime}(R)\ 
\sum_\nu Y_{\lambda\,\nu}(\hat{\bf R}) \ \ \mathcal{C}^{\sc (\lambda\nu)}_{n,n^\prime}\ ,
\label{mat-el-j1}
\end{equation}
where $I^{(\lambda)}_{n,n^\prime}(R)$ is the integral of Eq.~(\ref{intrad}) and
\begin{multline}
\mathcal{C}^{\sc (\lambda\nu)}_{n,n^\prime} = \sum_{m,m^\prime}
\left\langle l_n\, j\, m\,\left(M_n-m\right)\vert J_nM_n\right\rangle \ \mathcal{B}^{\sc (\lambda\nu)}_{l_n m,l_{n^\prime} m^\prime} \\
\times \left\langle J_{n^\prime} M_{n^\prime} \vert l_{n^\prime}\, j\,  m^\prime\, \left( M_{n^\prime} - m^\prime  \right) \right\rangle
\\\,
 \times \delta_{(M_n+m^\prime-M_{n^\prime} -m),0}\ 
 \delta_{(m+m^\prime+\nu),0}
\label{mat-el-j2}
\end{multline}

\medskip


%

\end{document}